\documentclass{article}
\pdfoutput = 1
\usepackage{graphicx}

\title{
A mathematical model of the discrete 3-disk \\
for the 3-dimensional Universe
}
\author{
Tetsuyuki Yukawa\thanks{e-mail address : yukawa@post.kek.jp}\\
{\it KEK theory center, Institute of Particle and Nuclear Studies} \\
{\it High Energy Accelerator Research Organization (KEK)}\\
{\it 1-1 Oho, Tsukuba, Ibaraki, 305-0801 Japan}
}
\begin{document}
\maketitle
\begin{abstract}
A mathematical model of the distribution function for the discrete 3-disk is proposed in order to utilize in the statistical evolution equation of the 3-dimensional Universe. 
The model is constructed based on analyses in known exact solutions of recursion equations for the generating functions of the discrete 2-disk.
The model distribution exhibits three types of phases characterized by geometrical structure of the disk with either 1, 2, or 3-dimensional nature.
Transitions between those phases are either cross-over, 1st order, or 2nd order depending on the model parameters.
The proposed distribution function is compared with numerical simulations of the dynamical triangulation(DT) with $ S^3 $, and $ D^3 $ topologies.
\end{abstract}
%%%%%%%%%%%%%%%%%%%%%%%%%%%%%%%%%%%%%%%%%%%%%%%%%%%%%%%%%%%%%%%%%%%%%%%%%%%%%%%%%%%%%%%%%%%%%%%%
%%%%%%%%%%%%%%%%%%%%%%%%%%%%%%%%%%%%%%%%%%%%%%%%%%%%%%%%%%%%%%%%%%%%%%%%%%%%%%%%%%%%%%%%%%%%%%%%
\section{Introduction}
%%%%%%%%%%%%%%%%%%%%%%%%%%%%%%%%%%%%%%%%%%%%%%%%%%%%%%%%%%%%%%%%%%%%%%%%%%%%%%%%%%%%%%%%%%%%%%%%
Discretization of space is one of the standard techniques for quantizing fields non-perturbatively.  
In case of the gravitational field a computational method, known as the dynamical triangulation(DT)\cite{DT2-KKM, DT2-ADF, DT2-D}, has been popularly tried up to four dimensions. 
In $2$-dimension numerical results have satisfactorily reproduced analytical expectations.
For an example, the string susceptibility predicted by the Liouville field theory\cite{LIOUVILLE-P,LIOUVILLE-D,LIOUVILLE-DK} was obtained for  $ S^2 $ and $ T^2 $ topologies with several central charges within statistical errors\cite{DT2DS-OTY}.
Great many numerical efforts have also been poured on higher dimensional cases\cite{Lattice}.
Those investigations have given some insights on the possibility of quantizing the gravitational field in four dimensions.
For an example, the grand-canonical DT simulation\cite{DT4DS-HEY} has shown the existence of second order phase transition when space is coupled to a certain number of matter fields.
However, no conclusive results were drawn at that time.
It may be not only because of lack of the computational ability but also lack of analytical supports for the method in contrast to the lattice QCD. 

While the discretization of space has been originally introduced as a technical tool to avoid the ultra-violet problem of quantizing gravity, there exists a possibility to regard the discrete space as the fundamental structure of space-time\cite{Amb,Loop,qc4}.
Based on this idea we have constructed the $d$-dimensional Universe as a complex of $d$-simplices with the $d$-disk topology, where the $ S^{d-1} $ boundary is assigned as space, while the direction perpendicular to the boundary surface as time.
Statistical evolution equation of the Universe has been formulated as the Markov process\cite{QU}, and applied to the $2$-point temperature correlation observed by the WMAP\cite{WMAP-DATA} in $4$-dimension.
In this model the evolution is regarded as a kind of the diffusion limited aggregation of simplices derived by entropic pressure.
For obtaining the distribution function of simplicial complexes we had to employ numerical simulations which have been the unique machinery beyond 2-dimension up to 4-dimension at that time.

As far as the 2-dimensional Universe is concerned mathematical studies of discrete geometry have shown significant results, such as giving analytic forms of distribution functions for various cases of discretization.
These results are elaborated mainly due to the graph counting theory originated by Tutte and his followers\cite{TUTTE,BROWNT,MULLIN}.
As an important application of the results the master equation for evolution of the 2-dimensional Universe has been solved analytically with using asymptotic limits of the exact distribution functions for the force field\cite{TY}.
However, in higher dimensions no analytic form of the distribution function was available, and all the essential properties were gathered only through numerical simulations, for examples\cite{DT3S-BK, DT3S-AM, DT3S-AV, DT3S-HTY, DT3D-WCR,DT3S-T, DT3S-RCK}, in the 3-dimensional case.\footnote[1]{Recently, the colored tensor model has been developed as an analytic model for higher dimensional counter part of the matrix model\cite{GURAU}.
We shall discuss relationships to our model in the last section.}

Recalling numerical efforts without sound theoretical support can conclude only a temporal consolation in the history of lattice quantum gravity, we focus the main aim of this paper to propose an analytical model of  distribution function for the discrete manifold.
We hope it will be able to provide a theoretical support for numerical methods and serve an initial input for the statistical evolution equation of the 3-dimensional Universe.
As one of the first trial we would like to propose an analytic distribution function for the discrete 3-disk which is constructed based on analyses of known exact results in two dimensions.

This paper is organized as follows.
In order to appeal the model to be reasonable, several distribution functions in 2-dimension are reviewed in the following section (Section 2).
In Section 3 a model distribution function for the discrete 3-disk is proposed and its asymptotic limit is calculated.
The predicted distribution function is applied for the $ S^3 $ and $ D^3 $ manifolds to compare with known numerical results obtained by DT simulations in Section 4.
The last section (Section 5) is devoted for discussions and comments on the evolution of the 3-dimensional universe and extension to the 4-dimensional manifold.
%

%%%%%%%%%%%%%%%%%%%%%%%%%%%%%%%%%%%%%%%%%%%%%%%%%%%%%%%%%%%%%%%%%%%%%%%%%%%%%%%%%%%%%%%%%%%%%%%
\section{Two dimensional distributions}
%%%%%%%%%%%%%%%%%%%%%%%%%%%%%%%%%%%%%%%%%%%%%%%%%%%%%%%%%%%%%%%%%%%%%%%%%%%%%%%%%%%%%%%%%%%%%%%
%%%%%%%%%%%%%%%%%%%%%%%%%%%%%%%%%%%%%%%%%%%%%%%%%%%%%%%%%%%%%%%%%%%%%%%%%%%%%%%%%%%%%%%%%%%%%%%
\subsection{analytical models of distribution functions}
%%%%%%%%%%%%%%%%%%%%%%%%%%%%%%%%%%%%%%%%%%%%%%%%%%%%%%%%%%%%%%%%%%%%%%%%%%%%%%%%%%%%%%%%%%%%%%%
Let us consider a triangulated 2-dimensional plane with the disk topology.
The triangulation is rooted by choosing a vertex on the boundary as the root vertex, and distinguishing the oriented edge on the boundary incident to the root vertex as the root edge.
The root edge is pointing under the right-hand screw rule directing upwards perpendicular to the plane.
Vertices and edges are called external if they lie on the boundary, and the remaining 1- and 2-simplices are internal.
The triangulation with $ n $ internal and $ m+3 $ external vertices is said to be of type $ [n,m] $.
Two triangulations $ T $ and $ T' $ is isomorphic when a one-to-one mapping exists from $ i $-cells($  i= 0,1,2 $) of $ T $ onto $ i $-cells of $ T' $, preserving incidence including root simplices.
Enumeration of the number of isomorphism classes of rooted triangulations has been initiated by Tutte\cite{TUTTE}, and subsequently refined by Brown\cite{BROWNT} and Mullin\cite{MULLIN} and many others for various cases of discretization. 

Tutte considered a class of triangulations in which no internal edge is incident with two external vertices ({\it 3-connected}).
The distribution of triangulations of type $ [n,m] $ is shown to be
%%%%%%%%%%%%%%%%%%%%%%%%%%%%%%%%%%%%%%
\begin{eqnarray}
f^T_{n,m} &=& 3 \sum_{j=0}^{min (m,n-1)} (m+j+2)(m-3j)  \nonumber \\
&\times& {4n+3m-j+1 \choose n-j-1,  3n+3m+3}{m+2 \choose j+1 ,  m-j+2}{m-1 \choose j  , m-j} \  (m \neq 0), \nonumber \\
f^T_{n,0} &=& 2 {4n+1 \choose  n+1 , 3n+2}, \label{eq.2.19}
\end{eqnarray}
%%%%%%%%%%%%%%%%%%%%%%%%%%%%%%%%%%%%%%
where the suffix $ T $ represents Tutte.
Here, we have defined the generalized binomial coefficient by
\begin{equation}
{n_1 \choose n_2 , n_3} = {n_1 ! \over n_2 ! n_3 !}. \label{gbc}
\end{equation}

Allowing two end vertices of an internal edge are external ({\it 2-connected})  Brown obtained the distribution function for type $ [n,m] $ to be
%%%%%%%%%%%%%%%%%%%%%%%%%%%%%%%%%%%%%%
\begin{equation}
f^B_{n,m} = 2 {2m+3 \choose m ,  m+2}{4n+2m+1 \choose n ,  3n+2m+3}, \label{eq.2.16}
\end{equation}
where the suffix $ B $ stands for Brown.
%%%%%%%%%%%%%%%%%%%%%%%%%%%%%%%%%%%%%%
These two classes of triangulations are required to satisfy that
(a) no two edges have the same end vertices, and
(b) every edge is incident with two distinct vertices.
Mullin considered a class of triangulations without the constraint (a) ({\it multiple-edges}), and obtain
%%%%%%%%%%%%%%%%%%%%%%%%%%%%%%%%%%%%%%
\begin{equation}
f^M_{n,m} = {2m + 3 \choose m+1,  m +1} 2^{n+1} {3n + 2m+2 \choose n ,  2n + 2m + 4}, \label{eq.2.15}
\end{equation}
where the suffix $ M $ represents Mullin.
%%%%%%%%%%%%%%%%%%%%%%%%%%%%%%%%%%%%%%
Recently, Krikun\cite{KRIKUN} considered the degenerate triangulations without the constraint (a) and (b) ({\it loops}).
The distribution function is shown to be
%%%%%%%%%%%%%%%%%%%%%%%%%%%%%%%%%%%%%%
\begin{equation}
f^K_{n,m}={2m+6 \choose m+2 , m+3} 2^{3n-4} {{3 \over 2} n+m+{1 \over 2} \choose n , {1 \over 2} n+m+{5 \over 2}},\label{eq.2.18}
\end{equation}
where the suffix $ K $ stands for Krikun. 
Brown\cite{BROWNQ} also considered the distribution function for the quadrangular dissection of type $ [n,m] $ with $ n $ internal and $ 2 m+4 $ external vertices.
The distribution function is given to be 
%%%%%%%%%%%%%%%%%%%%%%%%%%%%%%%%%%%%%%
\begin{equation}
g_{n,m} = 3 {3m+4 \choose m , 2m+3}{3n+3m+2 \choose n , 2n+3m+4}. \label{BQdst}
\end{equation}
%%%%%%%%%%%%%%%%%%%%%%%%%%%%%%%%%%%%%%

To our knowledge they exhaust all the known exact distribution functions of the 2-disk.
Analytical methods for extracting the distribution function from the generating function are often highly technical and sometimes intuitive, and we recommend readers to refer original papers for the details. 

It will be essential in our later construction of the distribution function for the discrete 3-disk to keep in mind that distribution functions of the Mullin, Brown, Krikun triangulation,  and the Brown quadrangulation have the product form of two generalized binomial coefficients.
The 2-dimensional manifolds considered in these four cases have a common feature that there exists a 1-dimensional sub-manifold embedded in the 2-dimensional plane.
By extracting $ n=0 $ components of distribution functions of triangulations, we find $ f^{B, M, K}_{0,m}=C_{m+1} $,
where $ C_{m} $ is the Catalan number, 
$$
C_m={2m \choose m,m+1},
$$
and the distribution function of quadrangle dissection gives
\begin{equation}
g_{0,m} =3{ 3m+2 \choose m \ 2m+3}. \label{1dsub}
\end{equation}
They all correspond to the distribution functions of branched trees made of either trivalent vertices for triangulations or tetravalent vertices for quadrangulation.
In contrast, the Tutte distribution function does not have such a simple product property, since it prohibits 1-dimensional sub-graphs.
We consider this is the main reason of the product nature of distribution function of maps which possesses the hierarchical structure.

%%%%%%%%%%%%%%%%%%%%%%%%%%%%%%%%%%%%%%%%%%%%%%%%%%%%%%%%%%%%%%%%%%%%%%%%%%%%%%%%%%%%%%%%%%%%%%%
\subsection{asymptotic distributions}
%%%%%%%%%%%%%%%%%%%%%%%%%%%%%%%%%%%%%%%%%%%%%%%%%%%%%%%%%%%%%%%%%%%%%%%%%%%%%%%%%%%%%%%%%%%%%%%
The asymptotic limit of a distribution function is achieved by taking the total number of 2-simplices large.
Writing ${\hat N_i}$ and ${\tilde N_i}$ for $ i=0,1 $ as total numbers of the internal and the external $ i-$simplices, respectively, we have the relationship either $ N_2 = 2{\hat N_0} + {\tilde N_1} -2 $ for triangulations, or $ N_2 = {\hat N_0} + {\tilde N_1}/2 -1 $ for quadrangulations.

Let us consider the asymptotic limit of a distribution function taking the Mullin distribution function as an example. 
Other cases are treated exactly in the same manner as this example except the Tutte distribution function, which will be considered separately. 
We introduce the new variables $ N = N_2 -1 $, and $ q = m/N $ together with $ p = n/N $.
Then, we have the relationship, $ p = {1 \over 2}(1-q) $, which implies the ranges of these variables as $ q= [0, 1] $, and $p= [0, 1/2] $.
By making use of the Stirling formula the generalized binomial coefficient has the approximate asymptotic form,
\begin{eqnarray}
 {a_1N + b_1 \choose a_2N+b_2, \  a_3N+b_3 } &\rightarrow &
{a_1^{b_1+1/2} \over a_2^{b_2+1/2} a_3^{b_3+1/2} } \left({a_1^{a_1} \over a_2^{a_2} a_3^{a_3}}\right)^N  N^{c-1/2}\\ \nonumber
& \times  & N^{(a_1-a_2-a_3) N} e^{-(a_1-a_2-a_3) N}, 
\label{afgbc}
\end{eqnarray}
where we write $ c = b_1-b_2-b_3 $, which will represent the critical exponent associating to a map whose distribution is expressed by the generalized binomial coefficient.
Applying the form to the Catalan number, where $a_1=2, a_2=a_3=1, b_1=b_2=0 $ and $ b_3=1 $, the asymptotic limit is written as
$$
C_N \rightarrow N^{-3/2} 2^{2 N}
$$
within an overall constant, where the exponent $-3/2 $ is characteristic to the 1-dimensional map.
We notice that the most diverging factor $ N^{(a_1-a_2-a_3) N} $ disappears in this case.
We shall observe that this diverging factor disappears in all the known distribution functions in the following analyses, and thus, the distribution is bounded exponentially.
  
The asymptotic limit of the Mullin distribution function expressed in terms of new variables $ (N,q) $ can be written as
\begin{equation}
f_N (q) \rightarrow N^{-2} q^{1/2}  \phi(q) \Phi(q)^N, \label{eq.2.22}
\end{equation} 
where 
\begin{eqnarray}
\phi(q) & = &{(3+q)^{{5/2}} \over (1-q)^{1/2} (1+q)^{9/2}}, \label{eq.2.23a}\\ 
\Phi(q) & = & 2^{-1/2} 2^{q/2} {(3+q)^{(3+q)/2} \over (1-q)^{(1-q)/2} (1+q)^{1+q}}. \label{eq.2.23b}
\end{eqnarray} 
In the asymptotic limit the distribution is dominated by the term $ \Phi(q)^N $.
Since the second derivative of the entropy density, which is defined by 
\begin{equation}
 s_N(q)= -{1 \over N} \log [f_N (q)], \label{entropyD}
\end{equation} 
is given asymptotically as
$$
{4 \over (1-q) (1+q) (3+q)},
$$
it is positive in the allowed range of $ q = [0,1] $, and the entropy density is a smooth concave function in this range.
The free-energy density defined by $ -s_N (q) + \mu q $, then, has a peak at $ q_0 $ in $ [0,1) $ for appropriate values of the parameter $ \mu=\mu_0 $, satisfying 
$$
\mu_0 = {d s_N(q) \over d q} \Big |_{q_0} ,
$$ 
$ \Phi(q)^N $ is well-approximated by the Gaussian function for large $ N $ as
\begin{equation}
 \Phi(q)^N  \sim  \Phi(q_0)^N \exp[-{1 \over 2} a_0 N (q-q_0)^2], \label{eq.2.25}
\end{equation} 
with 
$$
 a_0 = {d^2 s_N(q) \over d q^2} \Big |_{q_0} .
$$
We obtain the asymptotic limit of the Mullin distribution function as
\begin{equation}
f_{N,m} \rightarrow  m^{1 \over 2} N^{-{5 \over 2}}\exp(-{a_0 \over 2 N} m^2+\mu m-\lambda N) \label{adisM}
\end{equation} 
within a constant, where $ \mu=a_0 q_0 $, and $ \lambda= a_0 q_0^2/2-\log [\Phi(q_0)]$.
As for the $q$-dependence of the function $ \phi(q) $, which is regular in the range $ q=[0,1) $.
We approximate it by replacing $ q $ to its peak value at $ q_0 $, and absorb in the overall constant.
The exponent $-5/2 $ of the variable $ N $ is characteristic to the 2-dimensional map.

The asymptotic distribution functions of the Brown triangulation, the Brown quadrangulation, and  the Kulikun triangulation can be obtained precisely in the same form as eq.(\ref{adisM}) with the same exponent, but different non-universal constants $ (a_0, \mu, \lambda) $.

As we can expect from the non-product form of the Tutte distribution function, derivation of its asymptotic limit is a little involved comparing to the previous cases.
Nonetheless, the method is basically the same as others.
We replace the intermediate sum over $ j $ by the integral over $ x =  j/m $, regarding the distribution to be a continuous function of $ x $ assuming $ m (=q N) $ is large.
The asymptotic form of the Tutte distribution function is then expressed as
\begin{equation}
f^T_N (q) \rightarrow N^{-{3\over2}} \int_0^{x_{max}} dx \ \phi_T(x,q) \Phi_T(x,q)^N \label{eq.2.28}
\end{equation} 
within a constant, where 
\begin{eqnarray}
\phi_T(x,q) & = &{ \{2+(1-x) q \}^{3/2} \{1-(1+2 x) q\}^{1/2} \over (1+q)^{7/2} } \nonumber \\
& &\times  { (1+x) (1-3x) \over x^2 (1-x)^3} ,    \label{eq.2.29}
\end{eqnarray}
and where 
\begin{eqnarray}
\Phi_T(x,q) & = & {\{ 2+(1-x)q \}^{2+(1-x)q} \over [\{1-(1+2x)q\}/2]^{\{1-(1+2x)q\}/2} \{3 (1+q)/2 \}^{3(1+q)/2}}  \nonumber \\ 
 & &\times \{x^x (1-x)^{(1-x)}\}^{-2q}.  \label{eq.2.30}
\end{eqnarray} 
The upper value of the range of integration is given by
$$
x_{max} = \left\{\begin{array}{ll}
               1 & \mbox{for} \ 0 \leq q \leq {1\over 3} \\
               {1-q \over 2 q} & \mbox{for} \ {1\over 3} \leq q \leq 1 .
\end{array} \right. 
$$

We shall first make the Gaussian approximation for $ \Phi_T(x,q)^N $ as a function of $ x $, and obtain 
$$
f^T_{N,q} \rightarrow N^{-{3\over2}} \int_0^{x_{max}} dx \phi_T(x,q) \Phi_T(x_0,q)^N e^{ -{N \over 2} a_0 (q) (x-x_0)^2 }.
$$
Here, the peak position $ x_0 $ is determined by solving the equation 
$$ 
{\partial {\hat s}_T(x,q)\over\partial x} = 0 , 
$$
where we define the asymptotic entropy density as $ {\hat s}_T(x,q) = -\log[\Phi_T(x,q)] $, and we obtain $x_0= (1-q)/(3-q) $.
We also write 
$$ 
a_0(q)= {\partial^2 {\hat s}_T(x,q)\over\partial x^2} |_{x=x_0} ,
$$
which turns out to be $ (3-q)^3 (2-q) q/(1-q)^2/6 $.

Since $ x_0 $ is smaller than  $ x_{max} $ for any $ q $ within the range $ [0,1) $, the integration over $ x $ is dominated by the contribution around the peak $ x_0 $. 
After carrying out the integration, we repeat the Gaussian approximation for  $ \Phi_T(x_0(q),q)^N $as a function of $ q $, and obtain
$$
f^T_{N,q} \rightarrow N^{-2} q^{1\over2} { (2-q_0)^{1/2}\over  (1+q_0)^{7/2}  (3-q_0)^{1/2} } \Phi_T(x_0,q_0)^N  \exp\{- {N \over 2}A_0 (q-q_0)^2\},
$$
where $ q_0 $ is determined by the equation 
$$
 {\partial s_T(x_0(q),q)\over \partial q} = 0 , 
$$
which we solve numerically obtaining $q_0\sim0.2067$, and where we write
$$ 
A_0= {\partial^2 s_T(x_0(q),q)\over\partial q^2} |_{q=q_0}, 
$$
which is about $ 2.682 $.
Finally, the asymptotic distribution appears precisely in the same form as other cases:
\begin{equation}
f^T_{N,m} \rightarrow m^{1 \over 2} N^{-{5 \over 2}}\exp(-{A_0 \over 2 N} m^2+\mu m-\lambda N), \label{adisT}
\end{equation} 
with $ \mu=A_0 q_0 $, and $ \lambda= A_0 q_0^2/2-\log [\Phi_T(x_0(q_0),q_0)]$.

It is worth noticing that despite the apparent difference of the Tutte distribution function from others, all the distribution functions in asymptotic regions fall in the same universal form with the same critical exponents $ m^{1 \over 2} N^{-{5 \over 2}} $.
This suggests that even a simple product of generalized binomial coefficients, which are supposed to reflect the hierarchical structure of the manifold, can produce the universal form of the distribution function in the asymptotic limit.
We shall rely on this observation for obtaining the 3-dimensional distribution function in the next section.
%

%%%%%%%%%%%%%%%%%%%%%%%%%%%%%%%%%%%%%%%%%%%%%%%%%%%%%%%%%%%%%%%%%%%%%%%%%%%%%%%%%%%%%%%%%%%%%%%%
\section{A mathematical model of the distribution \\ function for the 3-disk}
%%%%%%%%%%%%%%%%%%%%%%%%%%%%%%%%%%%%%%%%%%%%%%%%%%%%%%%%%%%%%%%%%%%%%%%%%%%%%%%%%%%%%%%%%%%%%%%%
So far no recursion equation of the generating function is known for three dimensions, and we would like to propose a mathematical model inspired by the knowledge we have gathered in 2-dimensional discrete manifolds.
We shall construct the 3-disk with a simple $ S^2 $ boundary from the 3-sphere.
%%%%%%%%%%%%%%%%%%%%%%%%%%%%%%%%%%%%%%%%%%%%%%%%%%%%%%%%%%%%%%%%%%%%%%%%%%%%%%%%%%%%%%%%%%%%%%
\subsection{construction of the discrete 3-disk}
%%%%%%%%%%%%%%%%%%%%%%%%%%%%%%%%%%%%%%%%%%%%%%%%%%%%%%%%%%%%%%%%%%%%%%%%%%%%%%%%%%%%%%%%%%%%%%
The 3-dimensional sphere is prepared as a simplicial complex with the total numbers of $ i $-simplices being $ {\cal N}_i \ ( i = 0,1,2,3) $.
The Euler-Poincare relationship for the $ S^3 $ topology gives
\begin{equation}
{\cal N}_0-{\cal N}_1+{\cal N}_2 -{\cal N}_3= 0, \label{eq.3.0} 
\end{equation} 
together with the dimensional constraint,
\begin{equation}
{\cal N}_2 = 2 {\cal N}_3 \label{eq.3.1}.
\end{equation} 
The 3-dimensional disk may be obtained from the 3-sphere by choosing one vertex among $ {\cal N}_0 $, and removing all the tetrahedra sharing the vertex.
We call the removed complex as the cap, and the remaining disk as the target manifold.
They both have the $ D^3 $ topology with a common $ S^2 $ boundary.
The number of tetrahedra $ x $ of the cap is called the coordination number of the vertex.
Numbers of $ i $-simplices for the cap and the target manifold are written as $ \{ C_i \} $ and $ \{ N_i \} $, respectively.
For the target disk $ {N_i} $ are decomposed into sums of those belonging to the boundary $ \tilde{ N}_i $ and its complement $ \hat{N}_i $ satisfying
$$ 
N_i = \tilde{ N}_i +\hat{N}_i  \ (i = 0, 1, 2),
$$
which are often called the external and the internal $ i $-simplex, respectively.
Since the original 3-sphere is divided by the common $ S^2 $ boundary into two 3-disks, we have the relationships
\begin{eqnarray}
{\cal N}_i &=& N_i+C_i-{\tilde N}_i   \  (i=0,1,2), \nonumber  \\
{\cal N}_3 &=& N_3+C_3  \label{eq.3.2}.
\end{eqnarray} 
According to the construction procedure the cap variables are given to be
\begin{equation}
C_0={\tilde N}_0+1, \
C_1={\tilde N}_0+{\tilde N}_1, \
C_2={\tilde N}_1+{\tilde N}_2, \
C_3={\tilde N}_2. \label{eq.3.3}
\end{equation} 
By adding the two topological conditions of the $ S^2 $ boundary: 
\begin{equation}
{\tilde N}_0 -{\tilde N}_1+{\tilde N}_2=2, \
2 {\tilde N}_1= 3{\tilde N}_2 , \label{eq.3.4}
\end{equation} 
we count altogether 15 independent relationships among 18 variables.

Choosing $ {\hat N}_0, {\hat N}_1 $, and $ {\hat N}_2 $ as the independent variables, other variables representing the target 3-disk are expressed as
\begin{eqnarray}
N_0 &=& {\hat N}_2 - 2{\hat N}_1+3{\hat N}_0 + 4, \nonumber \\
 N_1 &=& 3{\hat N}_2 - 5{\hat N}_1+6{\hat N}_0 + 6, \nonumber \\
N_2 &=& 3{\hat N}_2 - 4{\hat N}_1+4{\hat N}_0 + 4 , \nonumber \\
 N_3 &=& {\hat N}_2 - {\hat N}_1+{\hat N}_0 + 1 , 
\label{ni}
\end{eqnarray} 
and
\begin{eqnarray}
{\tilde N}_0 &=& {\hat N}_2 - 2{\hat N}_1+2{\hat N}_0 + 4,  \nonumber \\
{\tilde N}_1 &=& 3{\hat N}_2 - 6{\hat N}_1+6{\hat N}_0 + 6 ,\nonumber \\ 
{\tilde N}_2 &=& 2 {\hat N}_2 - 4{\hat N}_1+4{\hat N}_0 + 4. 
\label{nitilde}
\end{eqnarray} 

 Let us consider the practical construction procedure of a possible 3-disk $ ({\hat N}_0,  {\hat N}_1 ,  {\hat N}_2) $.
We start with an elementary tetrahedron (0,0,0), and either attach or detach one tetrahedron to or from the target disk one by one just like a crystal growth through the diffusion limited aggregation.
%%%%%%%%%%%%%%%%%%%%%%%%%%%%%%%%%%%%%%%%%%%%%%
\begin{figure}
\begin{center}
\includegraphics[width=0.8\linewidth]{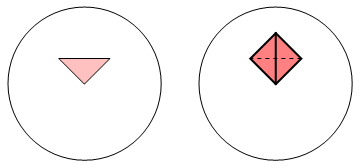}
\end{center}
\caption{$ {\cal M}_1 $-move \label{fig1}}
\end{figure}
%%%%%%%%%%%%%%%%%%%%%%%%%%%%%%%%%%%%%%%%%%%%%
\begin{figure}
\begin{center}
\includegraphics[width=0.8\linewidth]{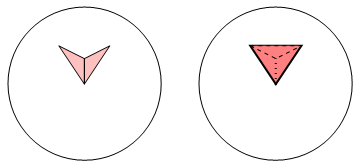}
\end{center}
\caption{$ {\cal M}_2 $-move \label{fig2}}
\end{figure}
%%%%%%%%%%%%%%%%%%%%%%%%%%%%%%%%%%%%%%%%%%%%%%%
\begin{figure}
\begin{center}
\includegraphics[width=0.8\linewidth]{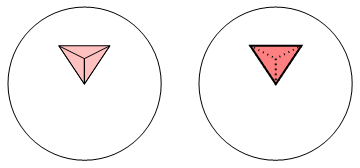}
\end{center}
\caption {${\cal M}_3 $-move \label{fig3}}
\end{figure}
%%%%%%%%%%%%%%%%%%%%%%%%%%%%%%%%%%%%%%%%%%%%%
There are 3 ways to attach a tetrahedron corresponding to the number of neighboring faces to be glued as shown in Figures 1, 2, and 3.
Each Figure shows a pair of circles. 
The left circle and the right circle represent the $ S^2 $ boundaries before and after gluing a tetrahedron, respectively.
Solid lines and dotted lines represent the external and the internal edges, respectively, while the dashed lines are also the external edges, which are hidden behind the attached tetrahedron. 
These gluing and removing procedures are called {\it moves}, which we represent as $ {\cal M}_i (i=1,2,3) $, and their inverses as $ \bar{\cal{ M}}_i $. They are sometimes called up-moves and down-moves, respectively.
Application of these three up-moves on a target 3-disk changes of variables $ \hat{ N}_i $, and $ N_i $, as listed in the following Table,
$$
\begin{array}{|c||c|c|c|c|c|c|c|c|}\hline
 moves &\Delta {\hat N}_2 & \Delta {\hat N}_1 & \Delta {\hat N}_0  & \Delta N_3 & \Delta N_2 & \Delta N_1 & \Delta N_0 & \Delta {\tilde N}_2 \\ \hline \hline
{\cal M}_1  &  1   &      0   &     0   &     1 &  3  & 3 &  1   &    2 \\ \hline
{\cal M}_2  &   2    &     1   &     0    &    1 &  2  & 1 &  0    &   0 \\ \hline
{\cal M}_3   &  3     &    3    &    1     &   1 &  1 &  0 &  0   &    -2 \\ \hline
\end{array}
$$
where we add $\Delta  {\tilde N}_2 $ in the list for the later use.

Let us first show that any possible 3-disk complex can be constructed by appropriate combinations of these three types of up and down moves.
Writing a net number of the ${\cal M}_i $-move as $ m_i $, {\it i.e.}number of  ${\cal M}_i $-moves minus number of  $ \bar{\cal{ M}}_i $-moves, the target 3-disk acquires the following numbers of simplices:
\begin{eqnarray}
N_0&=&4+ m_1, \nonumber \\
N_1&=&6+3 m_1+ m_2, \nonumber \\
N_2&=&4+3 m_1+2 m_2+m_3, \nonumber \\
N_3&=&1+m_1+m_2+m_3,\label{eq.3.7}
\end{eqnarray} 
and
\begin{eqnarray}
{\tilde N}_0&=&4+ m_1- m_3, \nonumber \\
{\tilde N}_1&=&6 +3 m_1- 3 m_3, \nonumber \\
{\tilde N}_2&=&4 +2 m_1 - 2 m_3,\label{eq.3.8}
\end{eqnarray} 
also
\begin{eqnarray}
{\hat N}_0&=& m_3, \nonumber \\
{\hat N}_1&=& m_2+3 m_3, \nonumber \\
{\hat N}_2&=& m_1+2 m_2+3 m_3. \label{eq.3.9}
\end{eqnarray} 
Inserting $ m_i (i=1,2,3) $ written in terms of $ {\hat N}_0, {\hat N}_1, {\hat N}_2 $ obtained from the last three equations:
$$
m_1={\hat N}_2 - 2{\hat N}_1+3{\hat N}_0, \
m_2 = {\hat N}_1 - 3 {\hat N}_0, \
m_3 = {\hat N}_0,
$$
the remaining 7 equations give the relationships, eq.(\ref{ni}) and (\ref{nitilde}).
This implies that the three types moves are sufficient to construct a simplicial complex with any possible number of variables, $ \{N_i, {\tilde N_i}, {\hat N}_i \} $.
However, this does not necessarily mean any possible complex can be constructed by some combinations of those 3 types of moves.

Secondly, let us show that it is indeed possible by searching a path starting from a certain possible complex, and choosing appropriate down moves to arrive at an elementary tetrahedron.
Then, the inverse path gives one of the actual construction of the desired complex.
We start with a $ D^3 $ complex ($  {\hat N}_0 , {\hat N}_1, {\hat N}_2$), which has $ {\tilde N}_2 $ external faces.
Each external face is sorted to the following 3 types:
a) one of 3 faces of a 3-simplex attached to the main body by sharing one face such as the complex created by an $ {\cal M}_1 $-move.
b) one of 2 faces of a 3-simplex glued to the main body sharing 2 faces such as that created by an $ {\cal M}_2 $-move. 
c) the face  of a 3-simplex glued to the main body sharing 3 faces such as that created by an $ {\cal M}_3 $-move. 
It is possible any time to remove the 3-simplex whose external face belong to the a-type.
It is not true when the face is either b- or c-type, in case when the down-move results non-allowed triangulation such as creating a pair of 3-simplices which are attached through a vertex or an edge, or creating a vertex or an edge whose neighboring simplices do not form the $ D^3 $, or the semi-$ D^3 $ when it is on the boundary.   

We begin with detaching all those simplices whose faces belong to the a-type.
Then the remaining external faces belong either to the b- or the c-type.
We select any one of simplices which are allowed to remove.
After the down-move we again check all the external faces and remove all those simplices whose faces belong to the a-type.
Repeating these procedures $ m_1+m_2+m_3 $ times we eventually arrive to an elementary tetrahedron.
Inverting these moves starting from a tetrahedron we can create the desired complex.
           
From the above construction procedures we recognize that there exist three hierarchical structures created by combinations of the 3 types of moves: i) applications of the $ {\cal M}_1 $-moves create complexes without any internal vertex and internal edge.
Then, all the internal triangles created by this type of moves have their associated three edges and vertices on the boundary like a branched tree, and thus these complexes belong to one-dimensional sub-manifolds.
ii) Combinations of the $ {\cal M}_1 $- and $ {\cal M}_2 $-moves create complexes having no internal vertices.
In such a complex two end-vertices of all the internal edges must lie on the boundary surface, that constrain the thickness of the 3-disks being the minimum.
It characterizes the manifold to be 2-dimensional like.
iii) Applying all the three types of moves, the resulting 3-disks are the 3-dimensional generic manifolds.

%%%%%%%%%%%%%%%%%%%%%%%%%%%%%%%%%%%%%%%%%%%%%%%%%%%%%%%%%%%%%%%%%%%%%%%%%%%%%%%%%%%%%%%%%%%%%%
\subsection{the model distribution function}
%%%%%%%%%%%%%%%%%%%%%%%%%%%%%%%%%%%%%%%%%%%%%%%%%%%%%%%%%%%%%%%%%%%%%%%%%%%%%%%%%%%%%%%%%%%%%%
We assume maps are rooted.
A rooted map of the 3-disk is obtained by specifying an external triangle as the rooted triangle, and selecting one of the three vertices and an arrowed edge pointing to the vertex under the right-hand screw rule directing outward as the rooted vertex and the rooted edge, respectively. 
It is sometimes more convenient and physical to choose $ {\tilde N}_2 $ as the independent variable rather than $ {\hat N}_2 $ through eq.(\ref{nitilde}), which we shall employ hereafter.

The distribution function of the rooted 3-disk is expressed by $ f_{n,l,m}^{D^3} $, where we define
\begin{equation}
n={\hat N}_0, \
l={\hat N}_1, \
m={1 \over 2} ({\tilde N}_2-4). \label{eq.3.16}
\end{equation} 
By removing a cap with the coordination number $ x $ from the $ S^3 $ manifold , we obtain the target disk with
\begin{eqnarray}
{\hat N}_0&=&{\cal N}_0-{x \over 2}-3, \nonumber \\
{\hat N}_1&=&{\cal N}_0+{\cal N}_3-2x-2, \nonumber  \\
{\tilde N}_2&=& x ,\label{eq.3.10}
\end{eqnarray} 
together with
\begin{equation}
N_3={\cal N}_3-x .\label{eq.3.11}
\end{equation} 
Therefore, once we obtain the coordination number distribution of the $ S^3 $ complex, $ P_{{\cal N}_0,{\cal N}_3}(x) $, we can calculate the distribution function for the target 3-disk, $ f^{D^3}_{n,l,m} $, as
\begin{equation}
f^{D^3}_{n,l,m}=P_{{\cal N}_0,{\cal N}_3}(x).
\label{eq.3.12}
\end{equation}

Corresponding to the three hierarchical structures there are three types of distribution functions.
When we apply only the $ {\cal M}_1 $-move on the manifold starting with a single tetrahedron, we can produce the restricted distribution function $ f_{0,0,m}^{D^3} $ corresponding to the 1-dimensional sub-manifold.
Maps created by the $ {\cal M}_1 $-moves are expected to be identical to the 2-dimensional quadrangular maps with zero internal vertex, which have the branched tree distribution function $ g_{0,m} $ of eq. (\ref{1dsub}), constructed by tetravalent vertices: 
$$
f_{0,0,m}^{D^3}= 3{3m+2 \choose m, \ 2m+3}.
$$
Here, we restrict ourselves considering only spacial degrees of freedom without any matter field coupled ({\it pure gravity}) for simplicity.

By including the $ {\cal M}_2 $-moves we shall obtain maps containing non-zero $ {\hat N}_1 $ and $ {\tilde N}_2 $. 
They give the distribution function of the form $ f_{0,l,m}^{D^3} $.
We expect from observations in the previous section that when the maps contain the 1-dimensional sub-manifold, the distribution function will be given by a product of two generalized binomial coefficients similar to $ g_{l,m}  $ eq.(\ref{BQdst}).
By making use of all types of moves we expect to construct a generic 3-dimensional manifold.

Now we give a naive but natural ansatz for the distribution function of the 3-disk, which possesses the double hierarchical structure, by a product of three generalized binomial coefficients, 
\begin{eqnarray}
f_{n,l,m}^{D^3} &=&X A^m {a_1m+a_2 \choose m, \ a_3 m+a_4}  B^l {b_1 m+b_2 l+b_3 \choose l, \ b_4 m+b_5 l+ b_6 } \nonumber  \\ 
& & \times \ C^n {c_1 m+c_2 l+c_3 n+c_4 \choose n, \ c_5 m+c_6 l+c_7 n+ c_8}. \label{eq.3.18}
\end{eqnarray} 
In this expression we count total 22 parameters, $A, B, C, X , \{a_i\}, \{b_i\}$, and $ \{c_i\}$, among which we are going to eliminate by imposing various conditions required either from theoretical or numerical reasons.
From the experience in the 2-dimensional exact distribution functions, the parameter sets $ {a_i, b_i, c_i} $
is expected to take either positive integer or half-integer values.
We consider here limiting within integer values for simplicity.

We should remind that the model distribution does not respect details of practical construction rules, such as the $ n $-conectedness, or acceptance of degenerate graphs.
Relying on the observations in the 2-dimensional exact distribution functions, we expect it is only in the asymptotic region the model distribution function reveals the universal form, which is free from constraints put in practical construction procedures.
 
One of the most basic theoretical requirement, which is not proven mathematically but seems to be true numerically, is the exponential boundedness of the number of independent geometries by the total number of 3-simplices $ N_3 $.
According to the asymptotic limit of the generalized binomial coefficients, the most diverging factor $ N^N $ is required to disappear.
It is satisfied by imposing three conditions,
\begin{eqnarray}
a_1-1-a_3+b_1-b_4+c_1-c_5=0,  \nonumber \\
b_2-1-b_5+c_2-c_6=0, \nonumber \\
c_3-1-c_7=0.
\label{condit1}
\end{eqnarray}

Hierarchical structure of  2-dimensional sub-manifold requires that the distribution function of the 2-dimensional sub-manifold
$$
f_{0,l,m}^{D^3} =X A^m {a_1m+a_2 \choose m, \ a_3 m+a_4} B^l {b_1 m+b_2 l+b_3 \choose l, \  b_4 m+b_5 l+ b_6 }{c_1 m+c_2 l+c_4 \choose 0, \ c_5 m+c_6 l+ c_8},
$$
should be written as a product of two generalized binomial coefficients.
It is achieved by asking
\begin{equation}
b_1=c_5, \ b_2=c_6, \ b_3=c_8,\label{condit2}
\end{equation} 
resulting
$$
f_{0,l,m}^{D^3} =X A^m {a_1m+a_2 \choose m, \ a_3 m+a_4} B^l {c_1 m+c_2 l+c_4 \choose l, \  b_4 m+b_5 l+ b_6 }.
$$
When no matter degree of freedom is involved, we ask the distribution function of 2-dimensional sub-manifold has the same critical exponent as the 2-disk distribution function:
\begin{equation}
c_4-b_6=-2. \label{condit3}
\end{equation} 

Similarly, the distribution function for the one dimensional sub-manifold,
$$
f_{0,0,m}^{D^3} =X A^m {a_1m+a_2 \choose m, \ a_3 m+a_4}{c_1 m+c_4 \choose 0, \  b_4 m+ b_6 },
$$
is required to be that for branched trees constructed by tetravalent vertices given by $ g_{0,m} $ eq.(\ref{1dsub}).
This gives 
\begin{equation}
a_1=b_4, \ a_2=b_6, \label{condit4}
\end{equation}
and
\begin{equation}
X=3, \ A=1, \ c_1=3, \ c_4=2, \  a_3=2, \ a_4=3. \label{condit5}
\end{equation}

At the end we obtain the 3-disk distribution function for pure gravity as
\begin{eqnarray}
f_{n,l,m}^{D^3} &=&3 {a_1m+4 \choose m, \ 2m+3} B^l {b_1 m+b_2 l+5 \choose l, \  a_1 m+b_5 l+4}  \nonumber \\
&\times & C^n {3 m+(b_5+1) l + c_3 n+2 \choose n, \ b_1 m+b_2 l+(c_3-1) n+ 5} , \label{modelD}
\end{eqnarray}
where the parameter $ b_3 $ is set to be $ 5 $, so that $ c_4-c_8=2-b_3 $ being $ -(manifold \\ dimension) $, which has been the case in  $ 1 $ and $ 2 $ dimensions.
In any way $ b_3 $ will not appear explicitly in the asymptotic limit.
We still count 7 free parameters at this moment left undetermined, which may be fixed by comparing to numerical simulations in the following sections.
%

%%%%%%%%%%%%%%%%%%%%%%%%%%%%%%%%%%%%%%%%%%%%%%%%%%%%%%%%%%%%%%%%%%%%%%%%%%%%%%%%%%%%%%%%%%%%%%%
\subsection{asymptotic distribution}
%%%%%%%%%%%%%%%%%%%%%%%%%%%%%%%%%%%%%%%%%%%%%%%%%%%%%%%%%%%%%%%%%%%%%%%%%%%%%%%%%%%%%%%%%%%%%%%%
In order to study the universal behavior of the distribution function which are expected to appear in asymptotic regions, we define new variables similar to the 2-dimensional cases corresponding to the independent variables $ n, l, m $:
\begin{equation}
p=n/N,  \ r=l/N, \ q=m/N, \label{eq.3.237}
\end{equation} 
where
\begin{equation}
N=N_3-1. \label{eq.3.28}
\end{equation} 
From eq.(\ref{ni}) and (\ref{nitilde}) there is a relationship,
\begin{equation}
q+r-p=1. \label{eq.3.29}
\end{equation}
We shall select $ (N,p,q) $ as the new set of independent variables when we discuss the asymptotic limit.

By making use of the asymptotic formula of the generalized binomial coefficient eq.(\ref{afgbc}), the model distribution function of the 3-disk in asymptotic regions, which we express as $ f^{D^3}_N (p,q) $, is written within an overall constant by
\begin{equation}
f^{D^3}_N (p, q) \sim N^{-5/2} q^{1/2} \phi (p, q) \Phi(p, q)^N, \label{3ddis}
\end{equation}
where
\begin{equation}
\phi(p, q)= { (b_1 q+b_2 r )^{11/2} \over r^{1/2} (a_1 q+b_5 r)^{9/2} } { (3 q+(b_5+1) r+c_3 p)^{5/2} \over p^{1/2} (b_1 q+b_2 r+(c_3-1) p)^{11/2}}, \label{eq.3.30b}
\end{equation} 
and where
\begin{eqnarray}
\Phi(p, q) & = & (a_1^{a_1}/4)^q B^r C^p { (b_1 q+b_2 r)^{b_1 q+b_2 r} \over r^r (a_1 q+b_5 r)^{a_1 q+b_5 r}} \nonumber \\ 
& \times & {\{3 q+(b_5+1) r+c_3 p \}^{3 q+(b_5+1) r+c_3 p} \over p^p \{b_1 q+b_2 r+(c_3-1) p\}^{ b_1 q+b_2 r+(c_3-1) p}} \label{3dPhi}.
\end{eqnarray} 
%

%%%%%%%%%%%%%%%%%%%%%%%%%%%%%%%%%%%%%%%%%%%%%%%%%%%%%%%%%%%%%%%%%%%%%%%%%%%%%%%%%%%%%%%%%%%%%%%
\section{ Phases of the stationary distributions \\ and comparison to numerical results}
%%%%%%%%%%%%%%%%%%%%%%%%%%%%%%%%%%%%%%%%%%%%%%%%%%%%%%%%%%%%%%%%%%%%%%%%%%%%%%%%%%%%%%%%%%%%%%%%
We now wish to check our proposition for the distribution function of the discrete 3-disk by comparing to numerical results.
Until now there are number of numerical studies available \cite{DT3S-BK, DT3S-AM,DT3S-AV, DT3S-HTY, DT3D-WCR,DT3S-T} for pure gravity with $ S^3 $ topology, and one result of $ D^3 $ \cite{DT3S-RCK}.
We shall look for appropriate values of parameters left undetermined in the last section from the numerical data.
%
%%%%%%%%%%%%%%%%%%%%%%%%%%%%%%%%%%%%%%%%%%%%%%%%%%%%%%%%%%%%%%%%%%%%%%%%%%%%%%%%%%%%%%%%%%%%%%%%
\subsection{$S^3$ topology}
%%%%%%%%%%%%%%%%%%%%%%%%%%%%%%%%%%%%%%%%%%%%%%%%%%%%%%%%%%%%%%%%%%%%%%%%%%%%%%%%%%%%%%%%%%%%%%%%
We can study statistical properties of the discrete $ S^3 $-manifold through the canonical partition function 
\begin{eqnarray}
Z_{{\cal N}_3}^{S^3}[\nu] &=& \sum_{{\cal T}(S^3)} e^{\nu {\cal N}_0}\\ \nonumber
&=& \sum_{{\cal N}_0} f^{S^3}_{{\cal N}_0,{\cal N}_3} e^{\nu {\cal N}_0},\label{eq.4.1}
\end{eqnarray}
where sums $ {\cal T}(S^3) $ are taken over all the independent triangulations with $ S^3 $ topology.
The distribution function $ f^{S^3}_{{\cal N}_0,{\cal N}_3}$ can be obtained from the $ D^3 $ distribution function with the least boundary area, which is created from the $ S^3 $ simplicial complex by removing one of $ {\cal N}_3 $ 3-simplices, as
\begin{equation}
 f^{S^3}_{{\cal N}_0,{\cal N}_3} = f^{D^3}_{n,l,0}/{\cal N}_3, \label{eq.4.2}
\end{equation}
where
$$
n={\cal N}_0-4,\ l={\cal N}_0+{\cal N}_3-6.
$$
In the asymptotic limit we get
\begin{equation}
 f^{S^3}_{{\cal N}_0,{\cal N}_3} \rightarrow f^{S^3}_{N}(p) \sim N^{-4} \hat{\phi} (p)  (BD)^N (BCD)^{pN} \hat{\Phi} (p)^N,\label{3sdis}
\end{equation}
where $ N={\cal N}_3-2, p=n/N $, and $ D=b_2^{b_2}/b_5^{b_5} $. 
After separating those terms to be included in either the normalization constant or the cosmological constants in $ \phi(p,0) $ and $ \Phi(p,0) $, we have
\begin{eqnarray}
\hat{\phi}(p) &\sim& {1 \over p^{1/2} }, \\ \nonumber
\hat{\Phi}(p) &\sim& { r^{(b_2-b_5-1) r} \over p^p} {\{(b_5+1) r+c_3 p\}^{(b_5+1) r+c_3 p} \over \{b_2 r+(c_3-1) p\}^{b_2 r+(c_3-1) p} }, \label{eq.4.3}
\end{eqnarray}
with $ r=1+p $.
We note the analytic part in $  \phi(p,0) $ is set to a constant because of  $ \hat{\Phi} (p)^N $ factor which dominates the sum over $ {\cal N}_0 $ in the large $ N $ limit. 

Asymptotically the partition function can be written in the integral form as
\begin{equation}
Z_{{\cal N}_3}^{S^3}[\nu] \sim N \int_0^{p_{max}} dp \ f^{S^3}_{N}(p) \ e^{\nu p N}.\label{eq.4.4}
\end{equation}
Upper value of the $ p $-integration $ p_{max}$ is determined  from eq.(\ref{eq.3.11}) in the limit $ N \rightarrow \infty $:  
\begin{eqnarray}
p &\rightarrow& {\eta-\xi/2 \over 1-\xi},\nonumber \\ 
q &\rightarrow& {\xi \over 2(1-\xi)},\label{rgpq}
\end{eqnarray}
where $ \xi=x/{\cal N}_3 $ and $\eta= {\cal N}_0 / {\cal N}_3 $.
Since the physical range of $ q $ is $ [0,1) $, $ \xi $ varies within the interval $ [0, 2/3) $, and $ p $ takes the value in $ (3 \eta-1,\eta] $ .
When  $ q \sim 1 \  ({\it i.e.} \ \xi \sim 2/3) $, most of the tetrahedrons have two neighbors to form a linear chain with its two faces attaching to neighboring tetrahedrons, while the remaining two faces left as external faces. 
This branched tree structure does not contain internal vertices, and the value $ p=0 $ at this limit fixes the maximum value of $ \eta $ to be $1/3 $, as conjectured by Walkup \cite{WALKUP}. 
The upper limit of $ p $ is fixed at the maximum value of $ \eta $:  $ p_{max} = 1/3 $.

One of the important findings of DT simulations is the nature of phase transition, which turned out to be of first order.
Property of the phase transition can be studied through the free energy density defined by
\begin{equation}
s_N^{S^3} (p) =  {1 \over N} \log [f_N^{S^3}(p)]. \label{eq.4.7}
\end{equation}
Asymptotically it converges to 
\begin{eqnarray}
s_N^{S^3} (p) &\rightarrow& \log[ {(1+b_5)^{1+b_5} \over b_2^{b_2}}] \ (p\rightarrow0),\nonumber \\ 
 &\rightarrow&  \log[ {(1+b_5+c_3)^{1+b_5+c_3} \over (b_2-1+c3)^{b_2-1+c_3}}] p \ (p \rightarrow \infty),\label{eq.4.12}
\end{eqnarray}
The second derivative of the free energy density,
\begin{equation}
s_{N}^{(2)}(p) \sim- {c_3 (c_3-1) (1+b_5-c_3) p + b_2 (1+b_5) c_2 r \over p (p+1) ((1+b_5) r + c_3 p)(b_2 r+(c_3-1) p)}, \label{2den}
\end{equation}
behaves as $-1/p $ for small $ p $, and as $ p $ gets larger it increases and cross zero once at $ p= p_1 $ with
\begin{equation}
p_1={1 \over  ({1 \over b_5+1}-{1 \over b_2}) c_3 (c_3-1) -1}.\label{eq.4.13}
\end{equation}
When $ p_1 $ lies below $ 1/3 $, the distribution function has two peaks, one at $1/3 $ and another between $ 0 $ and $ p_1 $, for an appropriate parameter range of $ \nu $.
In this case the system makes 1st order phase transition at the critical $ \nu $, where heights of the two peaks are equal.
Numerical simulation shows that the average $ N_0 $ has a jump at $ \nu=\nu_c \sim 4.0 $, with a small gap of $ \delta <p> \sim 0.03 $ \cite{DT3S-HTY}, and we shall choose the integer parameter triplet $ (b_2, b_5, c_3) $ in the region where $ p_1 $ is less than but close to $ 1/3 $.
We note that the system shows second order phase transition at $ p_1=1/3 $, and it becomes cross over transition as $ p_1 $ gets bigger than $ 1/3 $.
 
In order to fix parameters from the DT simulation results we need to make direct numerical integrations for the canonical partition function and other expectation values.
Among the unknown parameters there is no observable to determine the values directly except $ B $, and $ C $, which are related to the cosmological constants.
We shall fix the parameter triplet  $ (b_2, b_5, c_3) $ which control the nature of phase transition through $ p_1 $, by asking 1) they are positive integers and choose as small as possible just for simplicity, 2) they give $ p_1 $ reasonably close to $ 1/3 $ in order to make the jump in the average $ N_0 $ small, and 3) they give correct susceptibility peak pattern in the 2-dimensional parameter space of the $ D^3 $ topology, which will be discussed in the next sub-section.
At first we fix $ b_5 $ to be $ 3 $, after checking that $ b_5=1 $, and $ 2 $ do not satisfy the above condition 3.
With the fixed $ b_5 $ values of $ b_2 $ and $ c_3 $ are bounded below by the requirement $ p_1<1/3 $. 
Next, we choose $ b_2=6 $, which is the smallest possible value, and lastly $ c_3 $ to be $ 8 $ in order to make $ p_1 $ close to $ 1/3 $.
With these parameters the expectation value $ <p>(\sim {\cal N}_0/{\cal N}_3) $ and the finite size scaling of the $ {\cal N}_0 $-susceptibility are shown in Figures 4 and 5, respectively, which clearly exhibits the characteristics of first order phase transition.

%%%%%%%%%%%%%%%%%%%%%%%%%%%%%%%%%%%%%%%%%%%%%
\begin{figure}
\begin{center}
\includegraphics[width=.7\linewidth]{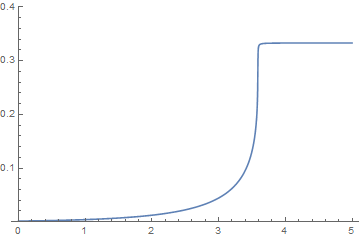}
\caption{$ <p>(\nu) $. $ N=100 $. $ \nu_c $ is fixed at $ 3.6 $.}
\end{center}
\label{fig4}
\end{figure}
%%%%%%%%%%%%%%%%%%%%%%%%%%%%%%%%%%%%%%%%%%%%%
\begin{figure}
\begin{center}
\includegraphics[width=.7\linewidth]{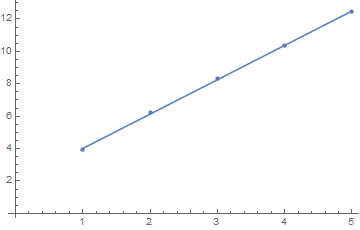}
\caption{ Finite size scaling of $\chi (\nu_N)$ for $ N=100,200,300,400,500 $. $\nu_N $ is chosen at the peak position of $ \chi(\nu) $.The line is for guiding eyes.}
\end{center}
\label{fig5}
\end{figure}
%%%%%%%%%%%%%%%%%%%%%%%%%%%%%%%%%%%%%%%%%%%%%
%

Two constants $ B $ and $ C $ are fixed from the position $ \nu_c $ of the jump in the average value $ < N_0 > $, and the asymptotic entropy density $ \mu_0 $ obtained from the canonical partition function eq.(\ref{eq.4.4}), which we parametrize as
\begin{equation}
{1 \over N} \log (Z_N^{S^3}[0]) \sim  \mu_0 -(3-\gamma) {\log N \over N} \label{eq.4.141} ,
\end{equation}
at a point $ \nu=0 $.
%%%%%%%%%%%%%%%%%%%%%%%%%%%%%%%%%%%%%%%%%%%%%%%
\begin{figure}
\begin{center}
\includegraphics[width=0.8\linewidth]{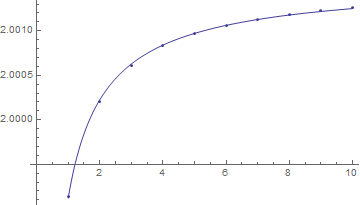}
\caption {Numerical fit of $\log (Z_N^{S^3}[\nu])/N $ by $\mu_0-(3-\gamma) {\log N \over N} $, with $ \mu_0=2.0 $ and $ \gamma=0.35 $ at $ \nu=0 $.}
\end{center}
\label{fig6}
\end{figure}
%%%%%%%%%%%%%%%%%%%%%%%%%%%%%%%%%%%%%%%%%%%%%
%
In Figure 6 we show $\log (Z_N^{S^3}[0])/N $ with a fixed $ \mu_0 \sim 2.0 $, which gives $ \gamma \sim 0.35 $. 
This parametrization differs from numerical simulations.
There were two numerical results available for the parameter $ \mu_0 $: in Ref.\cite{DT3S-AV} it was
$$
{1 \over N} \log (Z_N^{S^3}[0]) \sim  \mu_0 -\beta {1 \over N^{\alpha}} ,
$$
with $ \mu_0 \sim 2.06$, $ \alpha \sim 0.32 $ and $\beta \sim 3.90 $ , while in Ref.\cite{DT3S-AM} it was
$$
{1 \over N} \log (Z_N^{S^3}[0]) \sim  \mu_0 -\beta {1 \over \log N}
$$
with $ \mu_0 \sim 1.62$ and $ \beta \sim 1.44 $.
Our parametrization intends to give a meaning to the parameter $ \gamma $ to be the string susceptibility of the discrete $ S^3 $ manifold.
%%%%%%%%%%%%%%%%%%%%%%%%%%%%%%%%%%%%%%%%%%%%%
\begin{figure}
\begin{center}
\includegraphics[width=.7\linewidth]{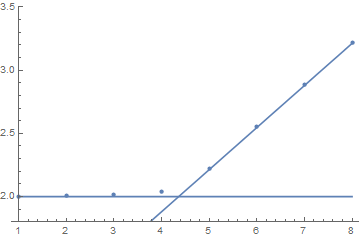}
\caption{$ \mu_0 (\nu)  $ Two lines are $\mu_0= 2 $, and $ 0.55+\nu/3$}
\label{fig7}
\end{center}
\end{figure}
%%%%%%%%%%%%%%%%%%%%%%%%%%%%%%%%%%%%%%%%%%%%%
\begin{figure}
\begin{center}
\includegraphics[width=.7\linewidth]{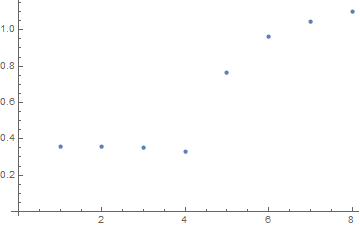}
\caption{  $\gamma (\nu) $}
\label{fig8}
\end{center}
\end{figure}
%%%%%%%%%%%%%%%%%%%%%%%%%%%%%%%%%%%%%%%%%%%%%
%
Numerically $ \mu_0 $ and $ \gamma $ may depend on $ \nu $.
In Figures 7 and 8 we show the $ \nu $ dependence of $ \mu_0 $ and $ \gamma $ obtained by numerical fitting of  $\log (Z_N^{S^3}[\nu])/N $ for $ N $ between $ 1000 $ and $ 10000$, respectively.
They are almost constant up until the transition point $ \nu \sim 4 $, and after that they behave differently.
This is because of the limit in the integration over $ p $ at $ p_{max} =1/3 $.
When $ \nu>\nu_c $, the peak at $ p=1/3 $ dominates in the integration.
As for the constant $ \nu_c $, there exist also two simulation results: $ \sim 4.0 $  determined from the $ S^3 $ DT-simulations\cite{DT3S-AV, DT3S-HTY}, and $ \sim 3.6 $ observed in the susceptibility map of the $ D^3 $ DT-simulation \cite{DT3D-WCR}.
These facts pretend us existence of the strong finite size effect and probably the systematic errors depending on numerical methods.

%%%%%%%%%%%%%%%%%%%%%%%%%%%%%%%%%%%%%%%%%%%%%%%%%%%%%%%%%%%%%%%%%%%%%%%%%%%%%%%%%%%%%%%%%%%%%%%%
\subsection{$ D^3 $ topology}
%%%%%%%%%%%%%%%%%%%%%%%%%%%%%%%%%%%%%%%%%%%%%%%%%%%%%%%%%%%%%%%%%%%%%%%%%%%%%%%%%%%%%%%%%%%%%%%%
Since the current study is intended for a construction of the 3-dimensional Universe, which stands alone in the non-framed 'space', there is no meaning to specify a root in the map, and the distribution should be counted for non-rooted maps.
Transformation of distribution functions from rooted to non-rooted $ D^3 $-maps is approximately performed through dividing the $ D^3 $-distribution function by  the number of possible root simplices, which equals to the number of boundary $2$-simplex $ {\tilde N}_{2}$.
If a map has spacial symmetries, such as rotational, reflectional, etc. this simple rule cannot be applied.
However, such symmetric maps are rare especially when $ N $ is large, and we expect they will not give much effects on the distribution function in the asymptotic limit.
We shall write the approximate distribution function of non-rooted maps as
\begin{equation}
{\hat f}_{n,l,m}^{D^3}={1 \over m} f_{n,l,m}^{D^3}. \label{eq.4.142}
\end{equation}

Statistical properties of the discrete $ D^3 $-manifold are studied through the canonical partition function 
\begin{equation}
Z_N^{D^3}[\mu,\nu] = \sum_{{\cal T}(D^3)} e^{-S[\mu,\nu]},\label{eq.4.151}
\end{equation}
where sums ${\cal T}(D^3)$ are taken over all the non-rooted maps with the $ D^3 $ topology.
Here the action for the $ D^3 $ topology is given by
\begin{eqnarray}
S[\mu, \nu]=\mu {\tilde N}_2 -\nu N_0 ,\label{eq.4.152}
\end{eqnarray}
In the asymptotic limit the partition function is written in the integral form as
\begin{equation}
Z_N^{D^3}[\mu,\nu] = N^2 \int_0^{1/3} dp \int_0^{q_{max}(p)} dq \ {\hat f}^{D^3}_N (p,q) e^{-\{(2 \mu-\nu) q-\nu p \} N},\label{eq.4.16}
\end{equation}
where $ p=n/N $ and $ q=m/N $ as before, and where
$$
{\hat f}_{N}^{D^3}(p,q)={\hat f}_{n,l,m}^{D^3}. \label{eq.4.17}
$$
The distribution function ${\hat f}^{D^3}_{N}(p,q)$ is defined from the original distribution function  ${\hat f}^{D^3}_{n,l,m}$by replacing $ l $ to $ N $ using the topological relationship $ N=l-n+m $.
For a fixed $ p $, eq.(\ref{rgpq}) gives $ q_{max}(p)=1-3p $.

The model distribution function for non-rooted $ D^3 $-maps in the large $ N $ limit is written as,
\begin{equation}
{\hat f}^{D^3}_N (p, q) \sim N^{-7/2} q^{-1/2} {\hat \phi} (p, q) B^N (BC)^{pN} (a_1^{a_1}/4/B)^{qN}   {\hat \Phi} (p, q)^N, \label{eq.4.17a}
\end{equation}
where
\begin{equation}
{\hat \phi}(p, q)= {1 \over r^{1/2} p^{1/2}},\label{eq.4.17b}
\end{equation}
and
\begin{equation}
{\hat \Phi}(p, q) = {q^{(a_1-3) q} \over r^r p^p} { (b_1 q+b_2 r)^{ b_1 q+b_2 r} \over  (a_1 q+b_5 r)^{ b_4 q+b_5 r}} {(3 q+(b_5+1) r+c_3 p)^{3 q+(b_5+1) r+c_3 p} \over (b_1 q+b_2 r+(c_3-1) p )^{ b_1 q+b_2 r+(c_3-1) p }}. \label{eq.4.76c}
\end{equation} 
In the above equations $ r $ should be replaced by $ 1+p-q $.
Analytic part of $ \phi(p,q) $ is set to a constant in the same reason as the $ S^3 $ case, and also in order to suppress the $ N $ dependence in numerical integrations with finite $ N $.
We need to make the double integrations for the patition function and other expectation values, which is time consuming.
For getting a better convergence and to save computer time we set $ N $ as small as possible, and choose $ N=30 $ in the following calculations for the susceptibility landscapes in the two parameter space.

To carry out the numerical integrations, we need to fix additional two integers for $ a_2 $ and $ b_1 $ for $ \hat{\Phi}(p,q) $.
Starting with $ a_1=1 $ and $ b_1=1 $, we find $ a_1 $ should be larger than $ 3 $ under the similar stuation as the condition on $ b_5 $.
After fixing $ a_1=3 $ we search appropriate $ b_1 $ starting from $ 1 $.
Until $ b_1 $ reaching to $ 5 $ the susceptibility peak range lying vertically is tilted, and we finally select $ (a_1, b_1)=(3,5) $.
We need to adjust the parameters $ B $ and $ C $ in order to fit the horizontal range of susceptibility peaks lying along $ \nu \sim 3.6 $, instead of the $ S^3 $-sumulation value of $ 4.0 $, and also $ \mu_c = 1.95 $, which lyes between two numerical results $ 1.6 $ and $ 2.0 $.
The agreement between the numerical result\cite{DT3D-WCR} and our model turned out to be unexpectedly good as we see in Figure 9 for the average $ N_0 $, and Figure 10 for the $ N_0 $-susceptibility.
%
%%%%%%%%%%%%%%%%%%%%%%%%%%%%%%%%%%%%%%%%%%%%%%%%%%%%%
\begin{figure}
\begin{center}
\includegraphics[width=.5\linewidth]{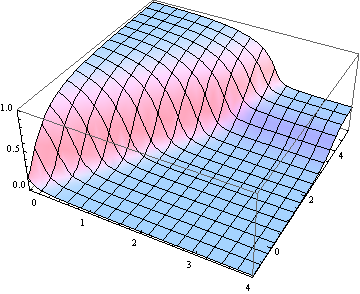}
\caption{$ \ <N_0>(\mu, \nu)/N $. $ N=30 $.}
\end{center}
\label{fig9}
\end{figure}
\begin{figure}
\begin{center}
\includegraphics[width=.5\linewidth]{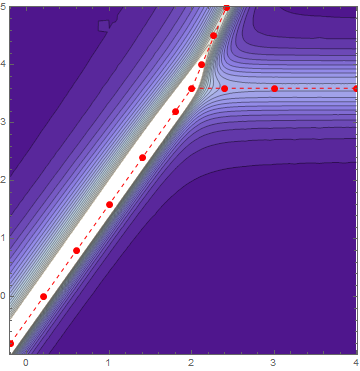}
\caption{$ N_0 $-susceptibility. Red dots are taken from \cite{DT3D-WCR} with dashhed lines for guiding eyes. }
\end{center}
\label{fig10}
\end{figure}
%%%%%%%%%%%%%%%%%%%%%%%%%%%%%%%%%%%%%%%%%%%%%%%%%%%%%%%%

We also calculate the average values of $ p \sim {\hat N}_0/N $, and $ q \sim {\tilde N}_2/N $ as shown in Figures 11 and 12, respectively.
They show clearly the existence of three phases reflecting the hierarchical structure of the $ D^3 $-manifold:
 i)  The platou in $ <p> \sim 1/3 $(blight area) correspond to the 3-dimensional generic manifold.
ii) The blight area in $ <q>\sim1 $ represents the 1D sub-manifold (branched tree) phase, and iii) the remaining area, which is the dark areas in both $ <p> $ and $ <q> $ is considered to be the 2-dimensional sub-manifold ($ <r> \sim 1 $).
These three areas are surrounded by susceptibility peaks determined by the DT-simulations.

%%%%%%%%%%%%%%%%%%%%%%%%%%%%%%%%%%%%%%%%%%%%%%%%%%%%%%%%
\begin{figure}
\begin{center} 
\includegraphics[width=.5\linewidth]{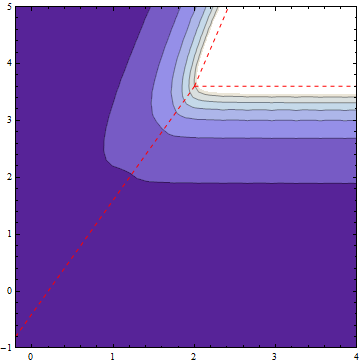}
\caption{$<p>(\mu,\nu)$. Hights of the blight platau is $ \sim 1/3$ and the dark bottom $\sim 0 $.}
\end{center}
\label{fig11}
\end{figure}
\begin{figure}
\begin{center}
\includegraphics[width=.5\linewidth]{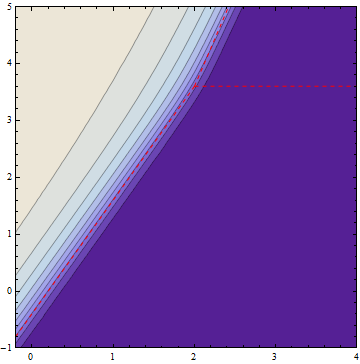}
\caption{$ <q>(\mu,\nu) $. Hights of the blight platau is $ \sim 1$ and the dark bottom $\sim 0 $. }
\end{center}
\label{fig12}
\end{figure}
%%%%%%%%%%%%%%%%%%%%%%%%%%%%%%%%%%%%%%%%%%%%%%%%%%%%%%%%%%%%%%%%%%%%%%%%%%%%%%%%%%%%%%%%%%%%%%%%
\section{Discussions and Summary}
%%%%%%%%%%%%%%%%%%%%%%%%%%%%%%%%%%%%%%%%%%%%%%%%%%%%%%%%%%%%%%%%%%%%%%%%%%%%%%%%%%%%%%%%%%%%%%%%
For predicting the distribution function of the discrete 3-disk we have made use of the common structural property in the exact forms of certain 2-dimensional disks.
We rely on the following two observations.
Firstly, when a discrete 2-disk manifold contains 1-dimensional sub-manifold the distribution function is written in a simple product of two generalized binomial coefficients.
Secondly, even if a manifold does not belong to above mentioned class, such as the Tutte distribution function, the distribution function will show asymptotically the same universal form with the same critical exponent as the product type.
In the 3-dimensional case the true distribution function of a manifold constructed by the DT may not be written in a simple product form.
However, we know that the discrete 3-disk constructed by the DT method possesses hierarchical substructure which persists in the asymptotic regions, and we expect a simple product distribution function of generalized binomial coefficients will show the same universal asymptotic behavior as that obtained by the DT.

In oder to check above prediction we compare our formula of the discrete 3-disk distribution function with numerical results obtained by the DT simulations.
We choose a set of parameters which fulfill exponential boundedness of the state volume, and show the hierarchical structure of sub-manifolds and 1st order phase transition, which is one of the significant feature of DT  simulation in $ S^3 $ manifold.
In general, our predicted distribution function shows the possibility of three types of transitions, either crossover, 1st order, or 2nd  order, depending on the choice of the parameter set.
Since 2nd order phase transition appears at a point on a line in the parameter space where phase changes from crossover to 1st order, we understand 1st order transition has more probability of appearance. 
There will be a possibility to obtain  2nd order phase transition if we put appropriate matter fields on 3-dimensional discrete space.  

Phase structure of the 3-dimensional manifolds is another important point to check.
It can be visualized by the expectation values of $ p $ and $ q $ in the two dimensional parameter space $ (\mu,\nu) $.
We observe three characteristic regions: i) for large $ \nu $ with small $ \mu $,  a plateau with the hight $ <p> \sim 1/3 $, which  corresponds to generic 3-dimensional manifold,
 ii) for large $ \mu $, a plateau with $ <q> \sim 1 $ which exhibits the 1-dimensional sub-manifold property,
 and iii) small $ \nu $ and $ \mu $ where $ <p>, <q> \sim 0 $, which is expected for the 2-dimensional sub-manifold.
Three phases are well-separated by two walls in the susceptibilities: one in $ \chi_p $ which separate the generic 3-dimensional disk and the 2-dimensional sub-manifold, and another in $ \chi_q $ which divides the 1-dimensional sub manifold from others.
Over all view of phase structure is quite similar to the numerical result \cite{DT3D-WCR}.

Our primary task of achieving an analytical form for the distribution function of 3-disk is to apply it for the statistical evolution equation of the 3-dimensional Universe.
The 'free energy' $ \log [f] $ acts as the force field in the master equation for the probability of the 3-dimensional Universe with the volume $ V=v_0 N_3 $, the surface area $ A=a_0 {\tilde N}_2 $, and the 'dilatation' $ D=d_0 {\hat N}_0 $ along the 'Markov-time' $ \tau $.
When we define the physical time $ t $ by
\begin{equation}
t=\int^{\tau} {1 \over A(\tau) } {dV(\tau) \over d\tau} d \tau,
\end{equation}
so that it satisfy the rate of change of 3-volume to be $ dV(t)=A(t) dt $, the relationship between the physical time and the Markov time is given by
$$
t \sim \left\{\begin{array}{ll}
               {1 \over 1-\kappa} \tau^{1-\kappa}  & \mbox{if} \ f\neq 1 \\
               \log \tau & \mbox{if} \ f= 1  
\end{array} \right. 
$$
for a solution of the master equation with the asymptotic behavior, $ V \sim \tau $ and $ A \sim \tau^\kappa $.
If we ask the physical time to posses the physical dimension $ 1 $, then the Markov time $\tau$ will have a dimension $ 1/(1-\kappa) $, and thus the body (V) and the face (A) carry the dimension $1/(1-\kappa)$ and $\kappa/(1-\kappa)$, for the case $ \kappa \neq 1$, respectively, while for the case $ \kappa=1 $ the space dimensions are infinity. 
We recall in the 2-dimensional Universe the statistical evolution equation had three asymptotic solutions, $ \kappa = (1, 1/2, 0) $, and thus the face dimension varies $ (\infty, 1, 0) $.
We expect for the 3-dimensional Universe there will be a $\kappa=2/3 $ solution in the master equation, so that the dimensions $ V $ and $ A $ are 3 and 2, respectively.
We are now underway to obtain solutions for three dimensional evolution equation.
We also focus our investigation on the effect of phase transition in the evolution of the Universe. 
Extension to the 4-dimensional case is straight forward, and it will soon appear in separate articles.

We would like to add some comments on the colored tensor model  which has been claimed to be a generalization of the matrix model for higher dimensions.
It is then regarded to be an analytic realization of the dynamical triangulation of space-dimensions higher than two.
In the model the dominant contributions come from graphs constructed with an elementary block called melon, which is formed by a pair of 3-simplices glued along 3 faces.
The generating function of rooted melon graphs is known to be 
$$
T(\lambda)=1+\lambda^2 T(\lambda)^4
$$
where $ \lambda $ is the vertex coupling constant in the dual graph of the 3-dimensional triangulation.
This equation can be interpreted as the generating function of branching trees made by 5-nary vertices, which is known to give the distribution function,
$$
T_m={4 m \choose m , 3 m +1},
$$
where $ m=(\tilde{N}_2 -4)/2 = N $.
Asymptotically the number of graphs is  $ T_N \sim N^{-3/2} e^{\mu_0 N} $ with $ \mu_0 =(1/2) \log(256/27) $, which is about $ 1.12 $.
On the other hand in the DT simulations it is either $ 1.6 $ or $ 2.0 $, which considerably surpasses the melonic graphs. 
We also remark that in the standard DT simulations the melon graph is excluded as degenerate graphs.
From the distribution functions $ T_m $ we can observe that the melonic graphs belong to the one dimensional universality class.
Recently, a new type of tensor model is introduced\cite{BENEDETTI-GURAU} with the critical property showing the 2-dimensional character in the double scaling limit.
We expect that another new tensor model which exhibit a genuine 3-dimensional phase transition will appear, so that the transition properties can be compared with our phenomenological model quantitatively.

%%%%%%%%%%%%%%%%%%%%%%%%%%%%%%%%%%%%%%%%%%%%%%%%%%%%%%%%%%%%%%%%%%%%%%%%%%%%%%%%%%%%%%%%%%%%%%%%

%%%%%%%%%%%%%%%%%%%%%%%%%%%%%%%%%%%%%%%%%%%%%%%% \end{document}
\end{document}